\documentclass[reprint,notitlepage,twocolumn,
superscriptaddress,aps,longbibliography]{revtex4-1}
\usepackage{graphicx} 
\usepackage{braket}
\usepackage{amsmath}
\usepackage{amsthm}
\usepackage{mathrsfs}
\usepackage{mathtools}
\usepackage{amssymb}
\usepackage{dsfont}
\usepackage{braket}
\usepackage{bm}
\usepackage{dsfont}
\usepackage{color}
\DeclareMathOperator{\tr}{tr}

\begin{document}
\title{Sequence-Model-Guided Measurement Selection for Quantum State Learning
\\
}

\author{Jiaxin Huang}
\affiliation{QICI Quantum Information and Computation Initiative, Department of Computer Science,
The University of Hong Kong, Pokfulam Road, Hong Kong}
\author{Yan Zhu}
\email{yzhu2@cs.hku.hk}
\affiliation{QICI Quantum Information and Computation Initiative, Department of Computer Science,
The University of Hong Kong, Pokfulam Road, Hong Kong}
\thanks{Jiaxin Huang and Yan Zhu contributed equally}
\author{Giulio Chiribella}
\email{giulio@cs.hku.hk}
\affiliation{QICI Quantum Information and Computation Initiative, Department of Computer Science,
The University of Hong Kong, Pokfulam Road, Hong Kong}
\affiliation{Department of Computer Science, Parks Road, Oxford, OX1 3QD, United Kingdom}
\affiliation{Perimeter Institute for Theoretical Physics, Waterloo, Ontario N2L 2Y5, Canada}
\author{Ya-Dong Wu}
\email{wuyadong301@sjtu.edu.cn}
\affiliation{John Hopcroft Center for Computer Science, Shanghai Jiao Tong University, Shanghai 200240, China}

\begin{abstract}
Characterization of quantum systems from experimental data is a central problem in quantum science and technology. But which measurements should be used to gather data in the first place? 
While optimal measurement choices can be worked out for small quantum systems, the optimization becomes intractable as the system size grows large.    
To address this problem,  we  introduce a deep neural network with a sequence model architecture  that  searches for   efficient  measurement choices in a data-driven, adaptive manner. The model can be applied to  a variety of tasks, including  the prediction of linear and nonlinear properties of quantum states, as well as state clustering and state tomography tasks.  
In all these tasks, we find that the  measurement choices identified by our neural network consistently outperform the  uniformly random choice.  
Intriguingly, for topological quantum systems, our model tends to recommend measurements at the system's boundaries, even when the task is to predict bulk properties. 
This behavior suggests that the neural network may have independently discovered a connection between boundaries and bulk, without having been provided    any  built-in knowledge of quantum physics. 
\end{abstract}

\maketitle

\section*{Introduction}
Machine learning provides a powerful tool for characterizing quantum systems based on measurement data~\cite{gebhart2023learning,huang2022provably,lewis2024improved,PhysRevResearch.6.033035,cho2024machine,sadoune2024learning,torlai2018,carrasquilla2019,tiunov2020experimental,PhysRevLett.127.140502,cha2021attention,zhong2022quantum,ma2023tomography,PhysRevResearch.6.033248,PhysRevResearch.6.023250,smith2021,zhang2021,wang2022,PhysRevB.107.075147,wu2023,koutny2023deep,tangtowards,PhysRevApplied.21.014037,PhysRevLett.132.220202,carrasquilla2017machine,van2017,liu2018,PhysRevE.99.062107,greplova2020unsupervised,PhysRevResearch.3.033052,jiang2023adversarial,dehghani2023neural,kim2024attention,iten2020,zhu2022,xiao2022,qian2023multimodal,wu2024learning,yao2024shadowgpt,huang2025direct}. 
 In particular, deep neural networks have played an important role across a range of tasks, including quantum state reconstruction~\cite{torlai2018,carrasquilla2019,tiunov2020experimental,PhysRevLett.127.140502,cha2021attention,smith2021,zhong2022quantum,ma2023tomography,PhysRevResearch.6.033248,PhysRevResearch.6.023250},    quantum similarity testing~\cite{zhang2021,wu2023,qian2023multimodal}, prediction of quantum entanglement~\cite{koutny2023deep,PhysRevLett.132.220202,huang2025direct}, and state classification~\cite{carrasquilla2017machine,van2017,liu2018,PhysRevE.99.062107,greplova2020unsupervised,PhysRevResearch.3.033052,jiang2023adversarial,dehghani2023neural,kim2024attention}. 
 Recent progress has enabled sequence models to predict diverse quantum properties of scalable quantum systems, by modeling the measurement outcome distributions~\cite{wang2022,PhysRevB.107.075147,tangtowards,PhysRevApplied.21.014037,fitzek2024rydberggpt,yao2024shadowgpt}.

An important question in quantum state learning is how to choose the appropriate measurements  to gather information about an unknown quantum state.  While an optimized adaptive choice can be found for small quantum systems~\cite{PhysRevLett.111.183601,PhysRevLett.113.190404,qi2017adaptive}, a full optimization quickly becomes intractable as the size of the system grows large. For scalable quantum systems,    
a widespread approach is to employ randomized measurements~\cite{brydges2019probing,huang2020,elben2020,elben2020(1),elben2020many,elben2023randomized,hu2025demonstration}.
This approach enables the estimation of a wide range of observables without performing a  full  tomography of the quantum state, which is not feasible  for large quantum systems. When prior knowledge is available, the randomized measurement choices can be further optimized~\cite{flammia2011,PhysRevLett.127.200503,PRXQuantum.5.010352}. In general, however, determining the optimal distributions is computationally challenging for large-scale quantum systems, especially when an approximated classical description is lacking.

In this work, we leverage a sequence model to automatically search for efficient  choices of measurement settings for a broad variety of quantum learning tasks.  Our network is based on a transformer architecture~\cite{vaswani2017attention}, which is designed to learn from sequential data.
We show that this network can be successfully used  to  predict properties of quantum many-body systems, to cluster quantum states based on their similarity, and  even to perform  a full  quantum state tomography for systems of small and intermediate size. The model dynamically adapts its measurement strategy based on previously collected data.   Unlike prior works on adaptive state tomography~\cite{quek2021,lange2023adaptive}, our approach is not limited to tomography tasks. Instead, it can be integrated as a subroutine in general quantum state learning workflows—enabling applications to large-scale quantum systems where full tomography is infeasible.


Our proposed method reduces the number of measurement settings required to achieve precise property prediction and accurate state tomography, compared with uniformly randomly sampling all possible measurements.
Moreover, it demonstrates superior performance in classifying quantum phases in an unsupervised manner with fewer measurements.
Our model demonstrates strong performance not only for states within the same class, but also for out-of-distribution quantum states, such as ground states of a perturbed Hamiltonian. 

Remarkably, when applied to topological quantum systems with open boundaries, our model discovers that detecting the system boundaries~\cite{gu09,PhysRevB.86.125441,pollmann12} can be significantly more effective than measuring the bulk directly for uncovering its quantum properties, even if those properties are functions of the bulk subsystem. 
This measurement preference emerges or vanishes depending on the presence of an open boundary, a behavior learned solely from local measurement data without prior knowledge of the system’s global configuration. 
This phenomenon highlights the model’s ability to uncover deep physical connections between boundary subsystems and bulk properties purely from data.  
These findings shed light on how AI models can automate complex quantum characterization experiments in the future. 

\section*{Results}
\subsection{Framework}

\begin{figure*}
    \centering
    \includegraphics[width=0.9\linewidth]{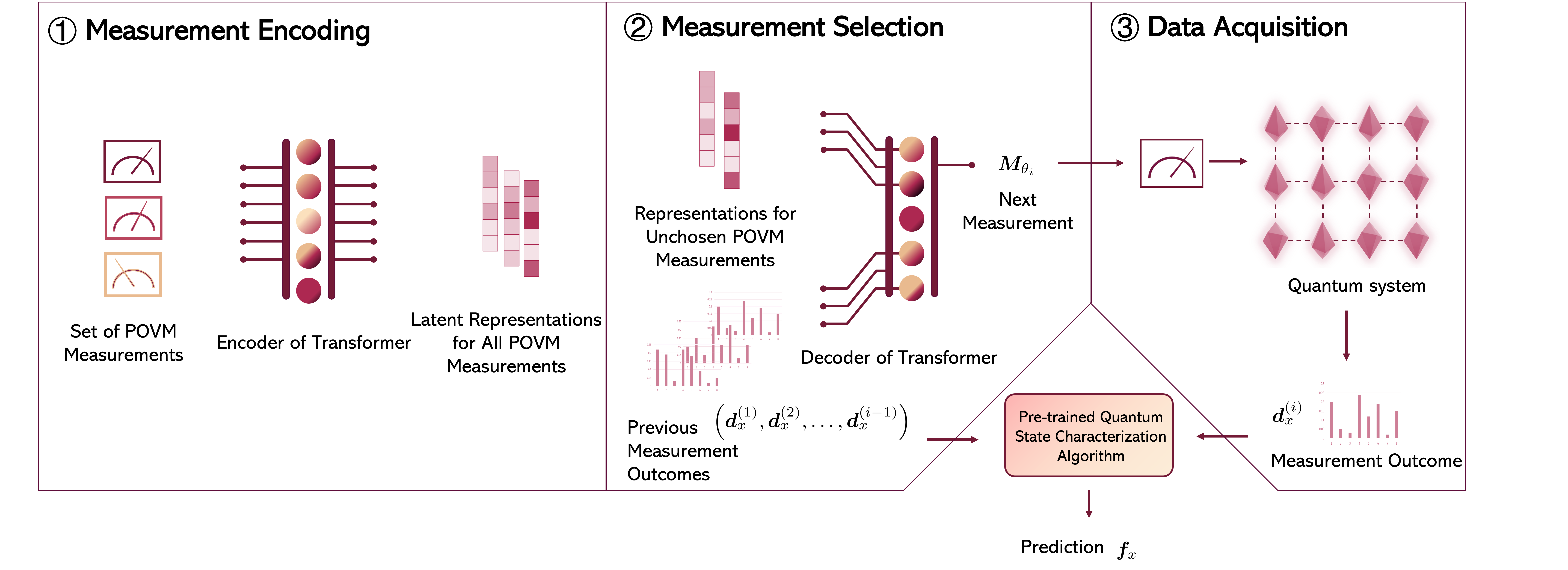}
    \caption{The learning procedure is divided into three stages: measurement encoding, measurement selection and data acquisition. In the measurement encoding stage, all possible POVM measurements are encoded into a latent representation space by a transformer encoder. In the measurement selection stage, a transformer decoders outputs the next POVM measurement based on the latent representations of all measurements and prior measurement data statistics. In the data acquisition stage, measurement data is collected with the POVM measurement selected by our model and then all measurement data are used to predict quantum properties. Throughout the learning process, the measurement encoding stage is performed first, followed by the alternating execution of the measurement selection and data acquisition stages.}
    \label{fig:model}
\end{figure*}

Here the aim is to learn a family of quantum states, for example, ground states $\ket{\psi(x)}$ of a certain parametrized Hamiltonian $H(x)$. Given a set of randomly sampled fiducial states $\{\ket{\psi(x)}\}_{x\in \chi}$, where $\chi$ is a set of physical parameters, a learner collects the measurement data $\bm d_x$ along with its associated label $\bm y_x$ for each state $\ket{\psi(x)}$. Each $\bm d_x$ consists of the outcome statistics $\left\{\bm d_x^{\theta}\right\}$ corresponding to a set of POVM measurements $\mathcal M:=\left\{\bm M_\theta:\theta\in\Theta\right\}$, where $\theta$ is a parametrization of measurements, $\bm M_\theta=\left\{M_j^\theta\right\}_j$ satisfies $M_j^\theta\ge 0$ and $\sum_j M_j^\theta=\mathbb I$. Ideally, if we ignore finite shot noise, then $\bm d_x^{\theta}=\{\braket{\psi(x)|M_j^\theta|\psi(x)}\}_{j}$; otherwise, each $\bm d_x^{\theta}$ is an approximation of the expectation values. The form of $\bm y_x$ depends on the task: for quantum property prediction, $\bm y_x$ can be a vector of quantum observable expectation values, while for quantum state tomography, $\bm y_x$ denotes density matrix $\ket{\psi(x)}\bra{\psi(x)}$.

We select an algorithm $f(\cdot)$ to produce predictions $\hat{\bm y}_x:=f\left(\bm d_x\right)$ of $\bm y_x$. This algorithm can be either a classical algorithm, like maximum likelihood estimation, or a well trained deep neural network model.  For each unknown state $\ket{\psi(x)}$, our transformer model adaptively produces a sequence of measurement parameters $(\theta^{(1)}, \theta^{(2)}, \dots, \theta^{(t)})$ based on the outcome data $\left(\bm d_x^{(1)}, \bm d_x^{(2)}, \dots, \bm d_x^{(t-1)}\right) (t\ge 2)$. 
During training, our transformer model is optimized to produce measurement sequences that, when combined with $f(\cdot)$, minimizes the prediction errors between $\hat{\bm y}_x$ and $\bm y_x$, averaged over $x\in \chi$.

After training is complete, the learner applies this trained model to adaptively produce measurement sequences for new quantum state $\ket{\psi(x')}$ unseen during training, based on collected measurement data. 
Specifically, as Fig.~\ref{fig:model} shows, the learning procedure is composed of three stages. The first stage is measurement encoding, where all possible POVM measurements from the set $\mathcal{M} = \{\bm M_\theta\}$ are embedded into a latent representation space using a transformer encoder. This encoding captures the structural and informational content of each measurement setting in a compact form, enabling the model to reason over the space of potential measurement strategies.

The second stage is measurement selection, where a transformer decoder sequentially selects the next measurement parameter $\theta^{(t)}$ based on the latent representations of all POVM measurements and the measurement history, i.e., the previously chosen measurements and their corresponding outcome statistics. This decoder models the conditional probability distribution
\begin{align*}
&\mathbb P\left( \theta \middle| \bm M_{\theta^{(1)}}, \bm{d}_x^{(1)}, \bm M_{\theta^{(2)}}, \bm{d}_x^{(2)}, \dots, \bm M_{\theta^{(t-1)}}, \bm{d}_x^{(t-1)} \right), \\
&\forall \theta\in\Theta\setminus\{\theta^{(1)},\theta^{(2)},\dots,\theta^{(t-1)}\}
\end{align*}
and outputs the measurement that is expected to be most informative for improving the prediction of $\bm y_x$.

The third stage is data acquisition, in which the selected measurement $\bm M_{\theta^{(t)}}$ is performed on the quantum state $\ket{\psi(x)}$, and the resulting outcome statistics $\bm d_x^{(t)}$ are collected. These data are appended to the growing dataset for that state. After the measurement, the full set of collected data $\left(\bm d_x^{(1)}, \bm d_x^{(2)}, \dots, \bm d_x^{(t)}\right) $ is passed to the prediction algorithm $f(\cdot)$, which outputs the estimated label $\hat{\bm y}_x$.
Throughout this process, the measurement encoding stage is performed once at the beginning, while the measurement selection and data acquisition stages are executed iteratively. This adaptive loop enables the model to dynamically refine its measurement strategy in response to the information gained during earlier rounds, leading to more efficient and accurate learning of quantum state properties or reconstructions.

From now on, we call this transformer-based model for selecting quantum measurements as transformer-guided-measurement-selection (TGMS) model. To benchmark the performance of our TGMS model, we compare it against a baseline approach of random sampling, where a POVM measurement is randomly sampled from $\mathcal M$ at each round of measurement. Then employing the same algorithm $f(\cdot)$ for predicting $\bm y_{x'}$, we can evaluate which approach yields a more efficient learning process.

\subsection{Quantum property prediction}
\subsubsection{Cluster-Ising model ground states}
\label{sec: cluster ising}
We apply our transformer-based model for predicting quantum properties, using the ground states of a one-dimensional quantum system with local interactions as a testbed.
Specifically, we first consider the ground states of a nine-qubit cluster-Ising model with open boundaries~\cite{smacchia2011}
\begin{equation}\label{eq:clusterIsing}
    H_{\text{cI}}^{\text{open}}=-\sum_{i=1}^{N-2}\sigma_i^z \sigma_{i+1}^x \sigma_{i+2}^z-h_1\sum_{i=1}^N \sigma_i^x -h_2\sum_{i=1}^{N-1}\sigma_i^x \sigma_{i+1}^x.
\end{equation}
The learner performs a particular type of measurements: Pauli basis measurements on each neighboring qubit triplet. This type of short-range correlation measurement strategy is important when experimentally feasible measurement settings are restricted and has been employed to characterize many-body quantum states~\cite{lanyon2017,PhysRevX.14.031035,li2025detecting}. Since reduced density matrices can be used as a probe for identifying topological phases~\cite{xu2025diagnosing}, as shown later, our strategy of measuring local subsystems in Pauli bases provides a feasible way to uncover topological features in scalable quantum systems. At each round of measurement, our transformer-based model adaptively selects a three-qubit subsystem to measure along with the Pauli basis for each qubit.

\begin{figure}
    \centering

    \includegraphics[width=0.95\linewidth]{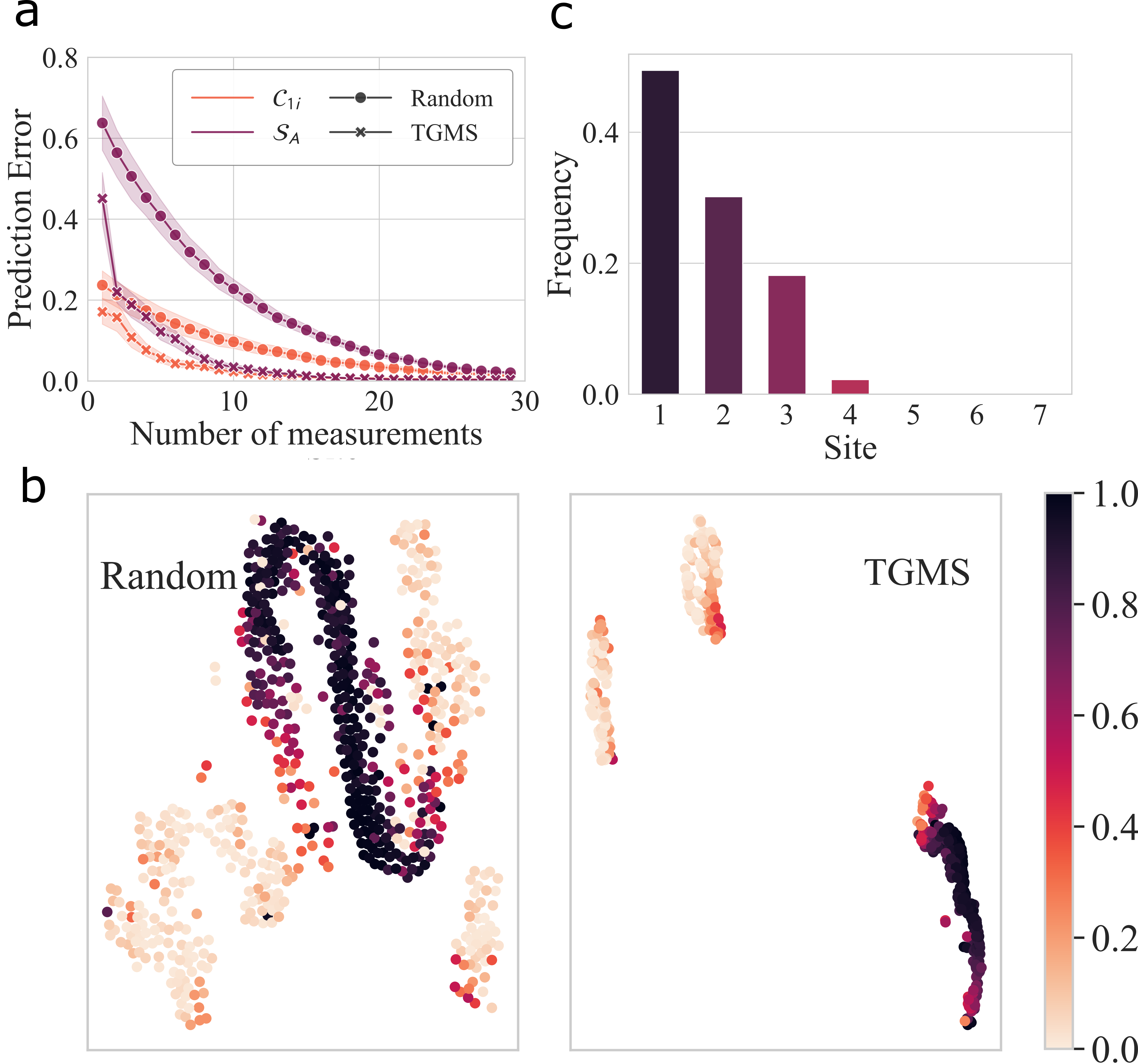}
    \caption{Prediction of properties for cluster Ising model ground states with chain configuration. Subfig a shows the prediction accuracies for both spin correlations and entanglement entropy with respect to the number of measurements. Subfig b shows the 2D projection of state representations (obtained by t-SNE algorithm) for both random sampling strategy and our learning model. Subfig c shows the frequencies of the Pauli measurements at different qubit triplets selected by our learning model, where the site index corresponds to the leftmost qubit of the triplet.}
    \label{fig:spinChain}
\end{figure}

 For property prediction, we employ a deep neural network, specifically a multi-task learning model~\cite{wu2024learning}. This model can not only predict properties including both spin correlations $\braket{\sigma_1^\alpha \sigma_j^\alpha}$, where $1\le j\le N$ and $\alpha\in\{x, z\}$, and entanglement entropies $-\log_2 (\tr\rho_A^2)$, where $A=[1,2,\dots, i]$ denotes a subsystem with $i$ spins, but also classify different phases of matter in an unsupervised manner. It is pre-trained over the dataset before being combined with our transformer-based model for inference.  
We optimize the transformer-based model to minimize prediction errors for both spin correlations and entanglement entropies. 
Once training is concluded, we deploy the TGMS model to adaptively select measurement settings for unseen cluster-Ising ground states. We compare this transformer-based model with the approach that randomly samples the set of all possible three-qubit-triplet Pauli measurements. The prediction accuracies illustrated in Fig.~\ref{fig:spinChain}{\bf a} implies that our transformer-based model achieves high prediction accuracy using significantly fewer measurement settings. Notably, our transformer-based model is flexible and can be combined with other deep learning models for property prediction.

Besides predicting spin correlations and entanglement entropies, the multi-task model generates state representations in latent space that enables unsupervised classification of different phases of matter~\cite{wu2024learning}. We project them onto a 2D plane using the t-SNE algorithm, where the color at each data point corresponds to the value of string order parameter~\cite{cong2019quantum,herrmann2022}
\begin{equation}\label{eq:stringOrder}
\braket{\tilde S}:=\braket{\sigma_1^z\sigma_2^x\sigma_4^x\dots\sigma_{N-1}^x\sigma_N^z}.
\end{equation}
For a fair comparison between our TGMS model and random sampling, we fix the number of measurement settings to $10$ in both strategies.
The results in Fig.~\ref{fig:spinChain}{\bf b} show that measurements selected by our TGMS model yield state representations that clearly separate into three distinct clusters, successfully distinguishing the symmetry-protected phase from the two symmetry-broken phases (paramagnetic and antiferromagnetic). In contrast, randomly sampled measurements yield adjacent representations that fail to distinguish these three phases.
This demonstrates our TGMS model's ability to identify quantum measurements that capture essential distinctions between different phases.

To understand this improved performance, we analyze the measurement patterns selected by our TGMS model. The results in Fig.~\ref{fig:spinChain}\textbf{c} reveals a strong preference for qubit-triplet measurements near a boundary of the quantum system. To demonstrate that the network’s boundary preference stems from patterns in the measurement statistics rather than qubit numbering, we test the model with permuted qubit indices.
This corresponds to the case, where it is unknown which qubits correspond to the boundary of a quantum system, for example, when the configuration of qubits in a many-body quantum system keeps changing. 
Under these conditions, the network consistently prioritizes measurements at system boundaries, confirming the above discovery. This finding is particularly interesting because while the conventional approach to distinguishing SPT phase from symmetry broken phase is to measure the entire string order parameter~(\ref{eq:stringOrder}), our model discovers a strategy of measuring spins at one boundary. This emerging phenomenon provides a fundamental insight: the critical information distinguishing topological states from trivial ones resides in their edge states. Consequently, when restricted to measuring short-range correlations, measuring the boundaries is more efficient than probing the bulk. Remarkably, this boundary-focused strategy emerges in a data-driven manner, without any explicit physical knowledge built into the neural network architecture.

\begin{figure}
    \centering

    \includegraphics[width=0.95\linewidth]{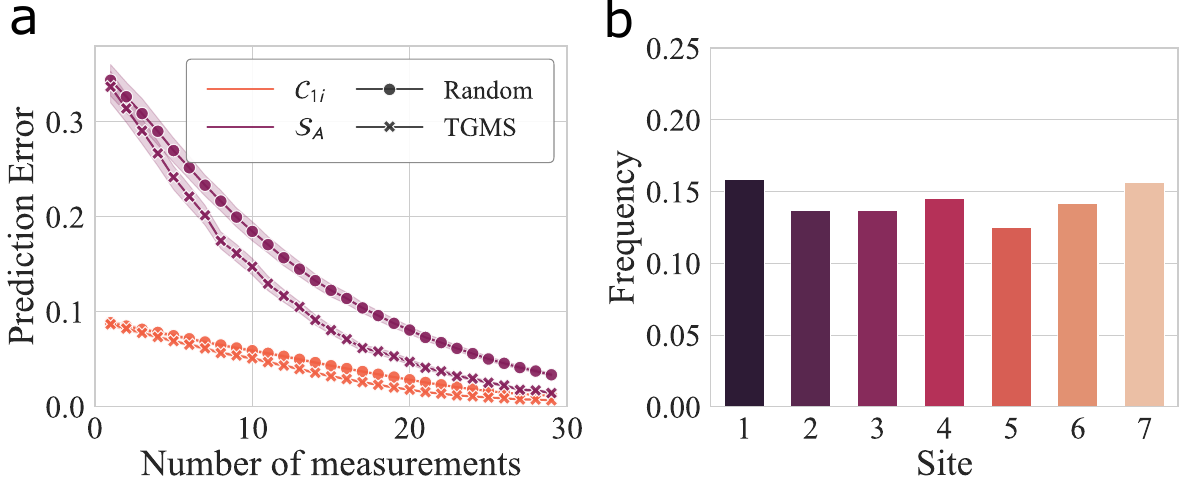}
    
    \caption{Prediction of properties for cluster Ising model ground states with ring configuration. SubFig a shows the prediction accuracies for both spin correlations and entanglement entropy with respect to the number of measurements. Subfig b shows the frequencies of the Pauli measurements at different qubit triplets selected by our learning model.}
    \label{fig:spinRing}
\end{figure}

To explore how the boundary conditions of quantum systems influence the measurement preference of our TGMS model, we consider the ground states of a cluster-Ising model with periodic boundaries, i.e.\ a spin ring configuration,
\begin{equation}\label{eq:clusterIsingRing}
    H_{\text{cI}}^{\text{peri}}=-\sum_{i=1}^{N}\sigma_i^z \sigma_{i+1}^x \sigma_{i+2}^z-h_1\sum_{i=1}^N \sigma_i^x -h_2\sum_{i=1}^{N}\sigma_i^x \sigma_{i+1}^x.
\end{equation}
where $N+1\equiv 1$ and $N+2\equiv 2$.
Applying the same learning scenario as before, we observe a fundamental difference in Fig~\ref{fig:spinRing}{\bf b}:  measurement selections become almost uniformly distributed across all three-qubit triplets, showing no spatial preference.
This striking contrast with open-boundary conditions demonstrates that our learning model has recognized the absence of boundaries within this periodic quantum system. These results suggest that our model can infer the global topology of a quantum system --- presence or absence of boundaries --- from its geometrically local measurement statistics in a data-driven manner. 

We also investigate the performance of our TGMS model using all-qubit measurements with periodic Pauli string bases of the form  $\sigma_1^i \sigma_2^j\sigma_3^k\sigma_4^i\sigma_5^j\sigma_6^k\cdots$ ($i,j,k\in \{x,y,z\}$), which has been experimentally implemented for characterizing many-body quantum systems~\cite{friis2018,karamlou2024probing}.  In this scenario, the neural network selects one from all possible $27$ measurements at each step. Our TGMS model outperforms the random sampling strategy again (the details can be found in Supplementary Material).

\begin{figure}
    \centering

    \includegraphics[width=0.9\linewidth]{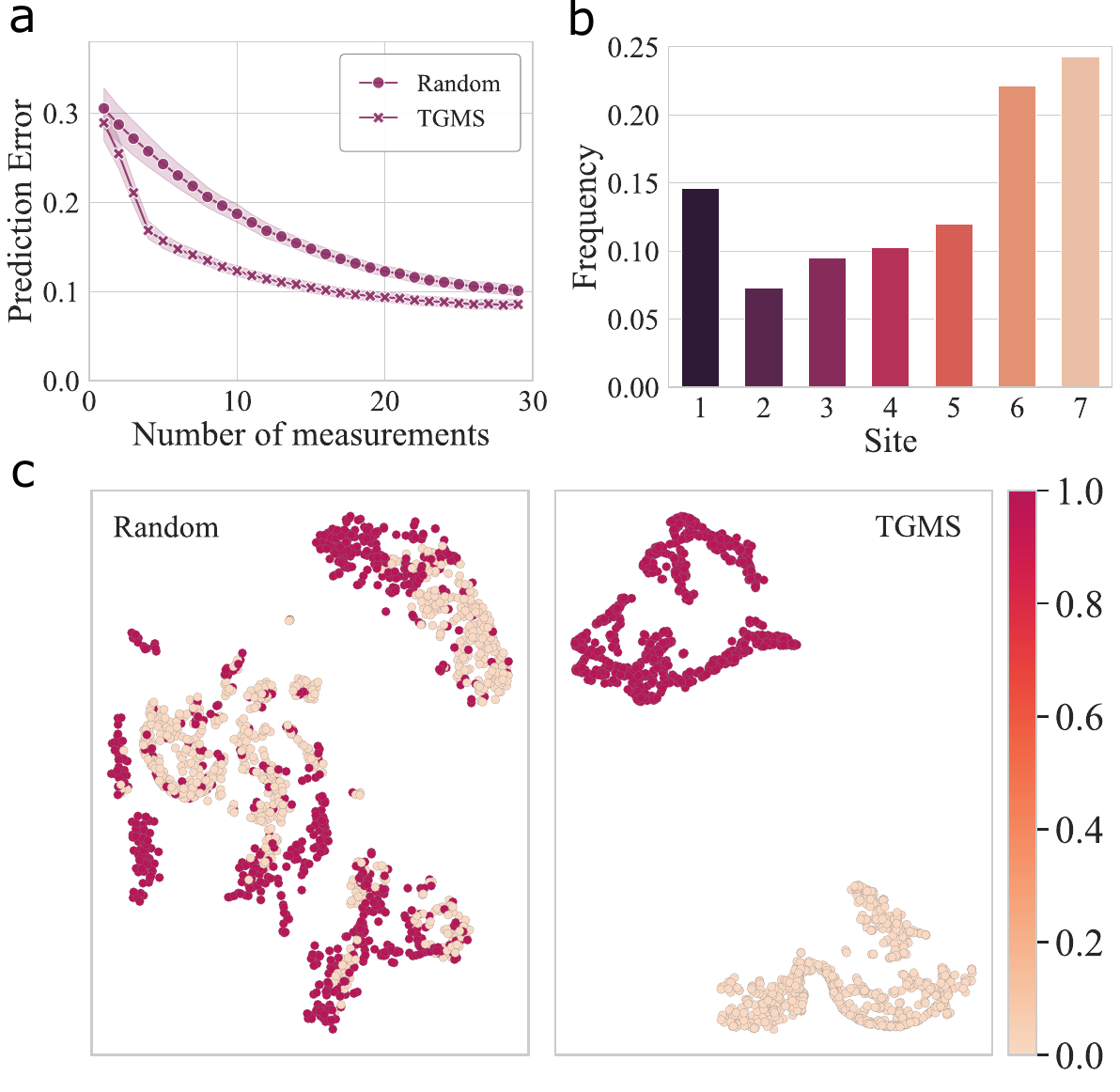}
    \caption{Prediction of properties for states generated by random quantum circuits without additional training. SubFig a shows the prediction accuracies for entanglement entropy with respect to the number of measurements. Subfig b shows the frequencies of the Pauli measurements at different qubit triplets selected by our learning model. Subfig c shows the 2D projection of state representations (obtained by t-SNE algorithm) for both random sampling strategy and our TGMS model.}
    \label{fig:randomGate}
\end{figure}

Note that measurements at different three-qubit subsystems can be performed simultaneously, hence reducing the required number of measurement settings. 
For this reason, we also investigate the performance of our adaptive algorithm using all-qubit measurements with periodic Pauli string bases of the form  $\sigma_1^i \sigma_2^j\sigma_3^k\sigma_4^i\sigma_5^j\sigma_6^k\cdots$ ($i,j,k\in \{x,y,z\}$). This type of periodic Pauli measurements has been experimentally implemented for characterizing many-body quantum systems~\cite{friis2018,karamlou2024probing} with a small number of measurement settings.  In this scenario, the neural network selects one from all possible $27$ measurements at each step. Our TGMS model outperforms the random sampling strategy again, achieving higher prediction accuracy for both spin correlations and entanglement entropies (details can be found in Supplementary Material). 

To further evaluate the generalization capability of our TGMS model, we test this trained model over out-of-distribution quantum states. 
Specifically, we aim at classifying phases of quantum states generated by shallow random circuit with $\mathbb Z_2\times \mathbb Z_2$ symmetric local gates. 
For shallow circuits, quantum states in both SPT and trivial phases remain within their respective phases after the application of random quantum circuits.
By employing our trained TGMS model to produce measurement sequences for these states produced by random circuits, we feed the resulting measurement data to our pre-trained neural network to obtain state representations.
Using the t-SNE algorithm, we find that the transformer-based model yields well separated clusters of state representations, as visualized by Fig.~\ref{fig:randomGate}{\bf c}, distinguishing between the SPT phase and the trivial phase (paramagnetic). In contrast, the distinction between these phases remains ambiguous when using randomly sampled measurements. 

This optimized measurement sequence produced by our TGMS model also yields more precise predictions of entanglement entropy, compared with the randomly sampled measurements, as demonstrated by Fig.~\ref{fig:randomGate}{\bf a}. To illustrate the patten of measurements selected our TGMS model, we plot the frequencies of measured qubit triplets (leftmost qubit) at different spin sites in Fig.~\ref{fig:randomGate}{\bf b}. The results show that the TGMS model retains a preference for boundary measurements due to the existence of SPT phase. However, since local random gates break the reflection symmetry with respect to the center of the spin chain, the original bias toward one boundary in Fig.~\ref{fig:spinChain}{\bf c} shifts to measuring both boundaries. Notably, this adaptation occurs autonomously, without additional training.

\subsubsection{Bond-alternating XXZ model ground states}
\label{sec: xxz}
Now we consider a learning task of a larger-sized quantum system.
Spcefically, we consider the ground states of a $50$-qubit bond-alternating XXZ model~\cite{elben2020many}
\begin{align}\notag  
H_{\text{XXZ}}=& J\sum_{i=1}^{N/2}\left(\sigma_{2i-1}^x \sigma_{2i}^x +\sigma_{2i-1}^y \sigma_{2i}^y+\delta\sigma_{2i-1}^z \sigma_{2i}^z\right)  \\   \label{eq:XXZmodel}
&+ J'\sum_{i=1}^{N/2-1}\left(\sigma_{2i}^x \sigma_{2i+1}^x +\sigma_{2i}^y \sigma_{2i+1}^y+\delta \sigma_{2i}^z \sigma_{2i+1}^z\right). 
\end{align}
There are three different phases, including topological SPT phase, trivial SPT phase and symmetry broken phase, in its ground state space.
 These three phases can be distinguished by a many-body topological invariant(MBTI)~\cite{elben2020many}
\begin{equation}\label{eq:TI}
\mathcal Z:=\frac{\tr(\rho_I u_I \rho_I^{T_1} u_T^\dagger)}{\left([\tr(\rho_{I_1}^2)+\tr(\rho_{I_2}^2)]/2\right)^{3/2}},
\end{equation}
where $\rho_{I_1}$, $\rho_{I_2}$ and $\rho_I$ are reduced density matrices on subsystems $I_1=[N/2- 9, N/2- 8, … , N/2]$, $I_2=[N/2 + 1, N/2+ 2, … , N/2+ 10]$ and $I=I_1\cup I_2$, $T_1$ denotes partial transpose on subsystem $I_1$, and $u_T:=\otimes_{i\in I_1}\sigma_i^y$.
MBTI in~(\ref{eq:TI}) is a nonlinear property of quantum state $\rho$, and hence cannot be directly estimated from measuring an observable. 

\begin{figure*}
    \centering

    \includegraphics[width=0.9\linewidth]{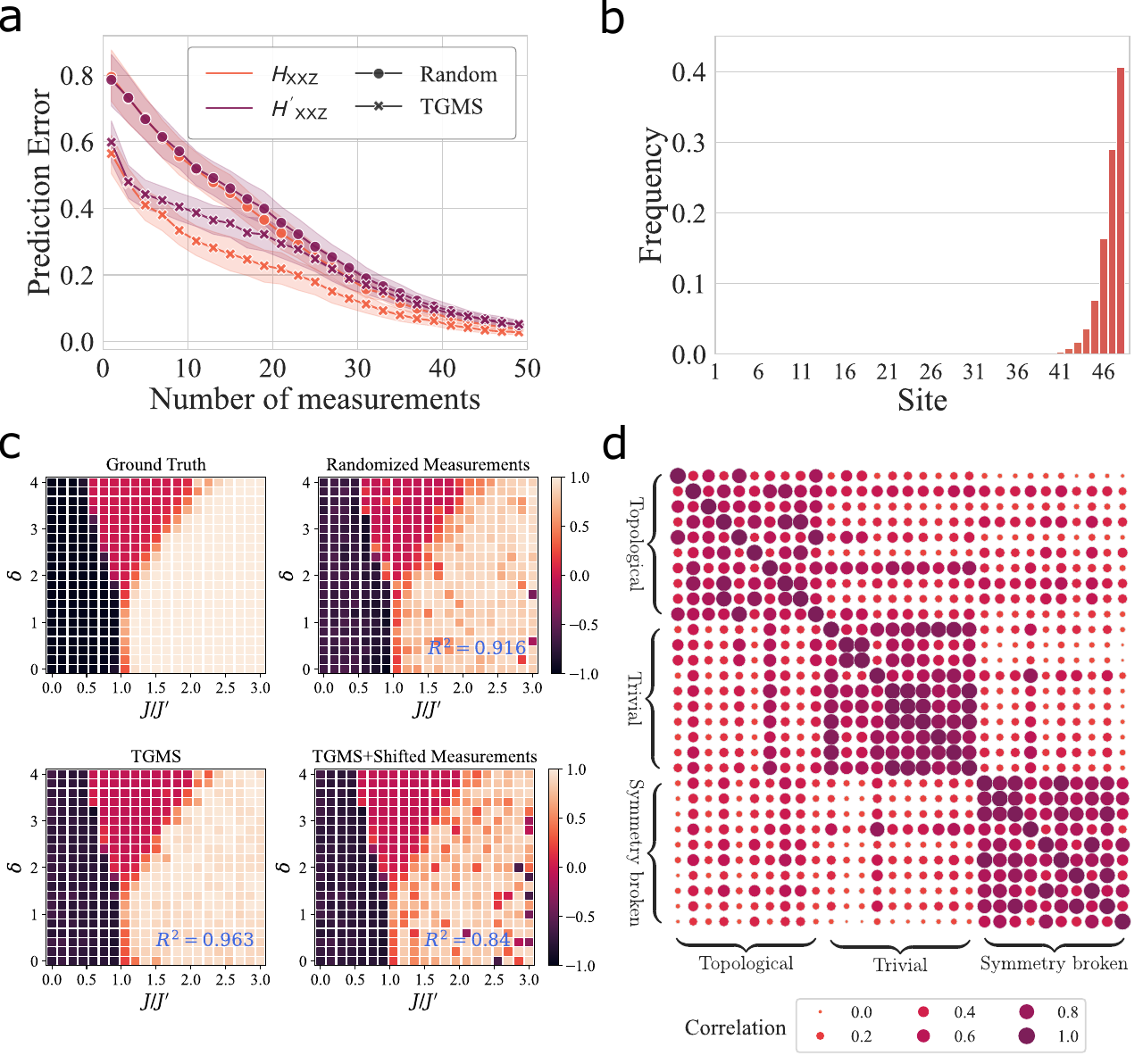}
    \caption{Prediction of properties for bond-alternating XXZ model ground states. Subfig a shows the prediction accuracies of MBTI values with respect to the number of measurements for ground states of $H_{\text{XXZ}}$, and the prediction accuracies of MBTI values for ground states of $H'_{\text{XXZ}}$ when our trained model is directly transferred to learn the ground states of $H'_{\text{XXZ}}$ . Subfig b shows the frequencies of the Pauli measurements at different qubit triplets selected by our learning model, where the site index corresponds to the leftmost qubit of the triplet. Subfig c shows the ground truth values of MBTI and predicted values of MBTI produced by three different measurement strategies: (M1) the Pauli measurements selected by our TGMS model, (M2) the Pauli measurements in (M1) but their sites are shifted towards left with $20$ qubits, and (M3) uniformly randomly sampled three-qubit Pauli measurements. Subfig d shows the state-wise correlation of measurement sequences produced by our TGMS model, where four ground states are randomly selected from each phase. Values closer to one corresponds to higher correlation while values closer to zero indicates lower correlations. }
    \label{fig:XXZ}
\end{figure*}

We pretrain a neural network to predict the MBTI from outcome statistics of Pauli measurement on neighbouring qubit triplet, following Ref.~\cite{wu2024learning}. We then train a transformer model to adaptively select three-qubit Pauli basis measurements.
 When tested on ground states of XXZ models not encountered during training, the TGMS model consistently favors measurements at a boundary of the quantum system. The frequencies of qubit-triplet measurements produced by our model at different sites averaged over all test states
 are illustrated in Fig.~\ref{fig:XXZ}{\bf b}. This measurement strategy outperforms random sampling of three-qubit Pauli measurements in prediction of MBTI, especially when the number of measurement settings is relatively small, as shown by Fig.~\ref{fig:XXZ}{\bf a}. 
 Remarkably, even though the MBTI (Eq.~(\ref{eq:TI})) is a nonlinear function of the reduced density matrix corresponding to a bulk subsystem, the TGMS model does not adopt the straightforward bulk measurement approach. Instead, a boundary-focused strategy emerges purely from data.   
 Similar findings are observed in a larger-sized bond-alternating XXZ model ground state with $100$ qubits (see Supplementary Material). 
 The above intriguing phenomenon suggests again that AI has the potential to uncover unconventional pathways for scientific discovery in physics.

To investigate how different measurement strategies affect the prediction accuracy for ground states in different phases, we compare the MBTI prediction accuracies for three strategies:
(\textit{M1}) measurement sequences generated by our model;
(\textit{M2}) Measurement sequences generated by our model and then shifted toward left by $20$ qubits;
(\textit{M3}) Uniformly randomly sampled measurement sequences.
The predicted MBTIs given by different measurement strategies at the entire phase diagram, together with the ground truth, are shown in Fig.~\ref{fig:XXZ}{\bf c}.

Comparing \textit{M1} and \textit{M2}, we find that shifted measurements yield worse prediction accuracy for states on the right side of the phase diagram, corresponding to the topological SPT phase. The reduced accuracy suggests that measuring system boundaries is crucial for topological quantum systems, as their edges encode essential bulk properties. 
In contrast, \textit{M1} and \textit{M2} provide almost exact same prediction accuracy for states in the trivial SPT phase. 
This suggests a key distinction between topological and trivial SPT phases: while information in topological phases is highly localized at system boundaries, trivial SPT phases exhibit approximately uniform information distribution across all local subsystems.
This finding naturally explains our model's inherent preference for boundary measurements --- it emerges from the need to accurately predict properties of topological SPT ground states.

To check whether our model employs distinct Pauli basis measurements for quantum states in different phases, we plot the correlations of Pauli strings selected by our model for ground states across three distinct phases of matter in Fig.~\ref{fig:XXZ}{\bf d}. The correlation between two measurement strategies is defined in \footnote{Here we neglect the spin sites and the order of  qubit-triplet Pauli measurement sequences. The measurements of two states are represented by two normalized $27$-dimensional vectors $f_i$ and $f_j$ after bin counting. Their correlation is defined as    $\operatorname{Corr}(f_i, f_j)\coloneqq \frac{\max_{i,j}\{d_{i,j}\} - d_{i,j}}{\max_{i,j}\{d_{i,j}\}-\min_{i,j}\{d_{i,j}\}}$, where $d_{i,j}=\lVert f_i-f_j\rVert_2$.}. The results demonstrate that for states within the same phase, the measurement settings chosen by our model exhibit relatively higher correlations, while for states from different phases, the correlations typically become lower.

Our model can be applied to produce measurement strategies not only for the ground states of the same class of Hamiltonian with different parameters, but also for the ground states of a perturbed Hamiltonian. To demonstrate this, we directly apply the trained TMGS model to produce the measurement sequences to predict the MBTI for the ground states of a perturbed Hamiltonian $H'_{\text{XXZ}}=H_{\text{XXZ}}+B\sum_{i=1}^N(\sigma_i^x\sigma_{i+1}^z-\sigma_i^z\sigma_{i+1}^x)$ with $B=0.1$. The results show that the TMGS model still achieves higher prediction accuracy than randomly sampled measurements, with similar performance for states before perturbation, indicating its strong generalization capability for out-of-distribution data.

\subsection{Quantum state tomography}
\label{sec: qst}

Now we test the performance of this TGMS model for tomography of quantum optical states, which are continuous-variable states on a quantized harmonic oscillator, Specifically, we consider quantum state reconstruction using sampled data of Husimi-Q function~\cite{PhysRevLett.120.090501,PhysRevLett.127.140502}.  
Given unknown state $\rho$, the learner can measure its Husimi-Q function 
\begin{equation}
Q_\rho(\alpha)=\braket{\alpha|\rho|\alpha}=\tr\left(\ket{0}\bra{0}D(-\alpha)\rho D(\alpha)\right),
\end{equation}
at any phase space point $\alpha\in \mathbb C$,
where $\ket{\alpha}$ is a coherent state, $\ket{0}$ is a vacuum state and $D(-\alpha)=D(\alpha)^\dagger$ is a displacement operator. This measurement can be realized~\cite{kirchmair2013observation} by applying displacement operation $D(-\alpha)$ on $\rho$ first and then performing a projective measurement $\{\ket{0}\bra{0}, \mathds{1}-\ket{0}\bra{0}\}$. 
After collecting sampled data of Q function at a set of phase space points $\{\alpha_i\}$ (a $16\times 16$ grid in the phase-space region $[-3, 3]\times [-3, 3]$), the learner obtains the dataset $\bm d:=\{d_i\}$, where each $d_i$ denotes the frequency of outcome $\ket{0}\bra{0}$ for $i$th phase space point. The learner can then estimate the density matrix of $\rho$ on a truncated Hilbert space (with dimension $d=16$)  from $\bm d$ by maximizing the likelihood $L(\tilde{\rho}|\bm d):=\prod_i \braket{\alpha_i|\tilde{\rho}|\alpha_i}^{d_i}$ over all possible states $\tilde{\rho}$. Specifically, we use iterative MLE algorithm~\cite{lvovsky2004iterative} for state reconstruction. Below, we neglect the statistical errors introduced by finite-number shots of measurements, and hence each data $d_i$ corresponds to the exact value of Q function at the sampled phase space point.

\begin{figure}
    \centering

    \includegraphics[width=0.9\linewidth]{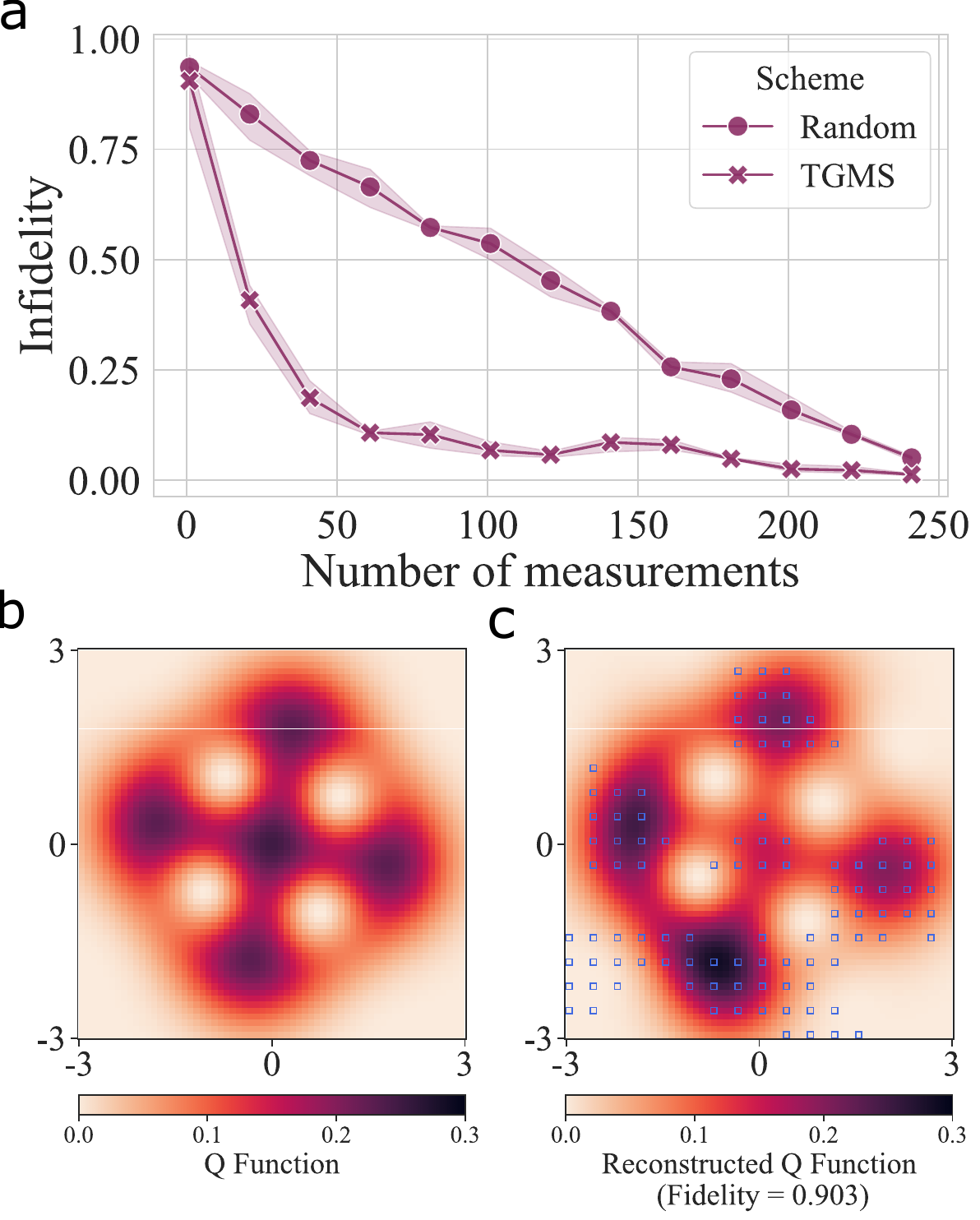}
    \caption{Tomography of cat states. Subfig {\bf a} shows infidelity of the reconstructed quantum states by random sampling measurements and by our TGMS model. Subfig {\bf b} shows the ground truth Q-function of an example state with $\alpha =-1.67-0.3 \text{i}$. Subfig {\bf c} shows the Q-function of the reconstructed state corresponding to the state in Subfig {\bf b} and the associated phase space points selected by our TGMS model. }
    \label{fig:catTomography}
\end{figure}

Inspired by the important application of cat states in bosonic quantum error correction~\cite{mirrahimi2014dynamically}, we consider the reconstruction of cat states with four coherent states in superposition.
We train our neural network model over a set of cat states $\frac{1}{\sqrt{2}}(\ket{0}_L+\ket{1}_L)$ with $\ket{0}_L:=\frac{1}{\sqrt{1+|\alpha|^2}}(\ket{\alpha}+\ket{-\alpha})$ and $\ket{1}_L:=\frac{1}{\sqrt{1+|\alpha|^2}}(\ket{\text{i}\alpha}+\ket{-\text{i}\alpha})$, where the amplitude of $\alpha$ falls within the region $|\alpha+1.5|\leq 1$.  During test, at $j$th step, using the collected data $\{Q_\rho(\alpha_i)\}_{i< j}$, the trained model produces the value of next phase space point $\alpha_j$ to be measured. The collected data of Q function at the sequence of phase space points are used to estimate the density matrix using the iMLE algorithm.
To benchmark the performance of our learning model, we compare it with the approach of measuring Q function over randomly sampled phase space points.  Note that conventional quantum state tomography, like MLE, requires at least $d^2-1=255$ sampled phase-space points to form an information complete measurement set for precise reconstruction. While the infidelity for random sampling decreases nearly linearly with the number of measurements,  our learning-based approach achieves a fidelity of around $0.95$ using only half the measurement data.

\section*{Discussions}
Transformer-based models have been extensively applied for quantum state tomography~\cite{cha2021attention,zhong2022quantum,ma2023tomography,PhysRevResearch.6.033248} as well as prediction of quantum properties across families of quantum states~\cite{wang2022,PhysRevB.107.075147,PhysRevApplied.21.014037,tangtowards,fitzek2024rydberggpt,kim2024attention,yao2024shadowgpt,suresh2025interpretable,rende2025foundation}. Unlike previous work using a transformer-based model to represent quantum states by predicting the sequences of their measurement outcomes, our TGMS model predicts the sequences of measurement settings for the task of property prediction and state reconstruction, enabling a more efficient learning process than randomly sampling measurement settings. 
This improvement stems from the neural network's ability to learn which measurement settings provide more information
about the unknown state. More discussion about related work can be found in the Supplementary Material.

For topological quantum systems with open boundaries, we find that our TGMS model has a priority to select measurements at the edges, while this preference disappears when the quantum system has periodic boundary conditions. It suggests that the neural network model has autonomously discovered that more information about the unknown state is concentrated at the edge of the quantum system and can effectively identify if such informative edge states exist based on geometrically local measurement data without any prior knowledge of its global topology. Remarkably, even when predicting the property determined by the reduced density matrix at a bulk subsystem, the TGMS model maintains a priority to measure the edges of the quantum system, uncovering the deep connection between edge states and bulk property in topological systems. 
During training on both topological and non-topological systems, our AI model gets this high-level understanding of the existence of edge states as a key difference between these two quantum phases, significantly improving its ability to distinguish between them.
The rediscovery of edge-bulk correspondence, together with the precise identification of this correspondence in topological systems in a data-driven manner, sheds light on how artificial intelligence can help physicists to uncover novel phenomenons in many-body quantum systems~\cite{iten2020,nautrup2022operationally,krenn2022scientific,wang2023scientific,wetzel2025interpretable} and automating quantum experiments~\cite{cao2024agents} in the future. 

Finally,  although this paper considers only one-dimensional quantum systems with SPT phases, this data-driven approach can be generalized to learning more complex quantum systems, like two-dimensional quantum systems with SPT phases and even quantum systems with topological order~\cite{xu2025diagnosing,PhysRevB.111.205125}.

\section*{Methods}
 In this section, we present the construction of dataset, together with the implementation details of the TGMS model and methods to train this model.
\subsection*{Data generation}

Here, we describe the data generation procedure for the training and test sets used in our model under different settings. In the quantum property prediction task, we exactly diagonalize the 9-qubit Hamiltonians of the cluster-Ising model to obtain ground states, which are then used to compute measurement outcome statistics, spin correlations, and entanglement entropies. For the bond-alternating XXZ model, we approximate the ground states and evaluate measurement statistics as well as MBTI values using the density-matrix renormalization group (DMRG) algorithm~\cite{schollwock2005density}. In the quantum state tomography task, the density matrices and Husimi-Q functions of cat states are generated using simulation tools provided by QuTiP~\cite{lambert2024qutip5quantumtoolbox}.

\subsection*{The TGMS model}

Here we present the neural network architecture of the TGMS model, which primarily consists of an encoder and a decoder. The encoder is responsible for measurement encoding phase, encoding POVM measurements into latent representations and the decoder is responsible for the measurement selection phase, outputting the next measurement based on collected data.

\subsubsection*{The encoder}

The encoder of the TGMS model encodes each POVM $\bm M_{\theta}$ as a $d_h$-dimensional latent vector $\bm h_\theta$, which will be used to represent $\bm M_\theta$ in all subsequent TGMS computation. The encoder architecture is similar to the Transformer encoder in \cite{vaswani2017attention}, but with no positional encoding. Thus the encoding results are invariant with respect to the input order:
$$\{\bm h_{\theta}\colon \theta\in\Theta\} = \operatorname{Enc}(\{\bm M_{\theta}\colon \theta\in\Theta\}). $$

Specifically, the TGMS encoder adopts a multi-layer structure. The first layer is a node-wise linear projection that unifies node dimensions to $d_h$. Subsequent layers are an interleaving of multi-head attention ($\operatorname{MHA}$) layers \cite{vaswani2017attention}, which executes inter-node information passing, and node-wise fully connected feed-forward ($\operatorname{FF}$) layers, which realizes self-update. The final encoding $\bm h_\theta$ of $\bm M_\theta$ is the output of the last layer. We further augment all layers by a skip-connection ($\oplus$) \cite{he2016deep} and a batch normalization ($\operatorname{BB}$) \cite{ioffe2015batch}. See Supplementary Material for detailed definitions of these encoding layers.

\subsubsection*{The decoder}
\newcommand{\thetainv}[1]{\Theta^{(#1)}}

At the $t$-th round of adaptive measurement selection, the decoder computes a probability
\begin{align}
\label{eq: P}
   \mathbb P\left( \theta \middle| \bm M_{\theta^{(1)}}, \bm{d}_x^{(1)}, \bm M_{\theta^{(2)}}, \bm{d}_x^{(2)}, \dots, \bm M_{\theta^{(t-1)}}, \bm{d}_x^{(t-1)} \right)
\end{align}
for all unused POVMs $\bm M_\theta$, with $ \theta\in \thetainv{t}\coloneqq\Theta\setminus\{\theta^{(1)},\dots,\theta^{(t-1)}\}$. The experimenter samples from this probability distribution to determine the next POVM to perform in the lab.

Specifically, when $t\geq 2$, the decoder first computes a $d_h$-dimensional latent representation $\bm h_x^{(t)}$ of quantum state $\ket{\psi(x)} $ using the measurement statistics $\left\{\bm M_{\theta^{(1)}}, \bm{d}_x^{(1)}, \bm M_{\theta^{(2)}}, \bm{d}_x^{(2)}, \dots, \bm M_{\theta^{(t-1)}}, \bm{d}_x^{(t-1)}\right\}$ gathered so far:
\begin{align*}
    \hat{\bm h}_x^{(i)} &= \operatorname{MLP}(\bm h_{\theta^{(i)}}, \bm d_x^{(i)}) ,\forall i\in \{1,\dots,t-1\},\\
     \bm h_x^{(t)} &=\frac{1}{t-1}\sum_{i=1}^{t-1} \hat{\bm h}_x^{(i)}.
\end{align*}
This latent vector $\bm h_x^{(t)}$ contains the information that the neural model has acquired about $\ket{\psi(x)}$. When $t=1$, no information has been collected. So the decoder uses a learned guess representation $\bm h_0^{(1)}$ independent of $x$.

Then, for each unused POVM $\bm M_\theta$, the decoder estimates its utility for further learning by assigning it a real-valued score:
$$u_\theta= \operatorname{Dec}(\bm h_x^{(t)}, \bm h_{\theta})\in \mathbb{R},\forall\theta\in\thetainv{t}.$$
Finally, the selection probability for each unused POVM is computed by applying a \textit{softmax} function on $u_\theta$ \cite{bridle1990probabilistic}.

The detailed formulation of the TGMS decoder is presented in Supplementary Material.

The decoder is optimized such that recurrently sampling from Equation \ref{eq: P} at each iteration maximizes the expected prediction accuracy. It is worth noting that in cases involving experimental constraints or cost considerations, the experimenter is allowed to manually select POVMs at certain iterations and resume sampling from Equation \ref{eq: P} in subsequent iterations.

\subsection*{Network training}
We split the dataset for each task into a training dataset, containing 80\% of total data, and a test set containing 20\% of data.

In the training phase,  in each measurement selection step $t$, we draw $K$ different samples $\{\theta_k^{(t)}\in \thetainv{t}:k=1,\dots,K\}$ from Equation \ref{eq: P}. Each $\theta_k^{(t)}$ induces an error $l_k^{(t)}=\lvert \bm y_x - f(\bm d_x^{(1)},\dots,\bm d_x^{(t-1)}, \bm d_{x, k}^{(t)}) \rvert^2$. We minimizes the weighted total error $L=\sum_{t=T_1}^{T_2}\sum_{k=1}^{K} u_{\theta_k^{(t)}}l_k^{(t)}$ over all $k$ and $t$. In fact, when $\theta_k^{(t)}$ has been chosen, $l_k^{(t)}$ is a constant with respect to TGMS parameters. So the above minimization effectively increases the utility $u_{\theta_k^{(t)}}$ for $\theta_k^{(t)}$ that corresponds to lower loss, and decreases the utility $u_{\theta_k^{(t)}}$ for $\theta_k^{(t)}$ that corresponds to higher loss. Therefore, when training is completed, a sample $\theta^{(t)}$ drawn from Equation \ref{eq: P} will likely produce a small prediction error. 

To improve training efficiency, we only count measurements selected within the sliding window $t\in[T_1, T_2]$ towards the total error. We progressively increase $T_1$ and $T_2$ during training, while fixing the window size $T_2-T_1$ unchanged. This trains the model to produce measurement sequences of arbitrary length using a constant gradient propagation depth.

\subsection*{Hardware}
The TGMS model is implemented using the PyTorch library and trained on one NVIDIA H800 SXM GPU.

\section*{Acknowledgment}
We thank the stimulating discussions with Mingpu Qin (SJTU), Chun Chen (SJTU), and Yuxuan Du (NTU). This work is supported by the Ministry of Science and Technology of China (MOST2030) under Grant No. 2023200300600. YDW acknowledges funding from the National Natural Science Foundation of China through grants no.\ 12405022. G.C. acknowledges support from the Hong Kong Research Grant Council through Grants No. 17307520, No. R7035-21F, and No. T45-406/23-R, by the Ministry of Science and Technology through Grant No. 2023ZD0300600, and by the John Templeton Foundation through Grant 62312, The Quantum Information Structure of Spacetime. The opinions expressed in this publication are those of the authors and do not necessarily reflect the views of the John Templeton Foundation. Research at the Perimeter Institute is supported by the Government of Canada through the Department of Innovation, Science and Economic Development Canada and by the Province of Ontario through the Ministry of Research, Innovation and Science.

\bibliography{refs}

\end{document}